\documentclass[prl,  longbibliography, reprint, 
amsmath,amssymb,amsfonts,floatfix]{revtex4-1}
\usepackage{braket}
\usepackage{graphicx}
\usepackage[hypertexnames=false,naturalnames=true]{hyperref}
\hypersetup{colorlinks=True}
\usepackage[all]{hypcap}
\usepackage{subfigure}
\usepackage[usenames,dvipsnames]{xcolor}
\usepackage{commath}
\usepackage{bbm}
\DeclareMathOperator{\Tr}{Tr}
\usepackage{times}

\usepackage{filecontents}

\begin{document}
\title{Heralded Bell State of Dissipative Qubits Using Classical Light in a Waveguide}
\author{Xin H. H. Zhang}
\email{xin.z@duke.edu}
\author{Harold U. Baranger}
\email{baranger@phy.duke.edu}
\affiliation{Department of Physics, Duke University, P.O.\,Box 90305, Durham, NC 27708-0305, USA}
\date{April 8, 2019}
\begin{abstract}

Maximally entangled two-qubit states (Bell states) are of central importance in quantum technologies. We show that heralded generation of a maximally entangled state of two intrinsically open qubits can be realized in a one-dimensional (1d) system through strong coherent driving and continuous monitoring.  In contrast to the natural idea that dissipation leads to decoherence and so destroys quantum effects, continuous measurement and strong interference in our 1d system generate a pure state with perfect quantum correlation between the two open qubits. Though the steady state is a trivial product state which has zero coherence or concurrence,  we show that, with carefully tuned parameters, a Bell state can be generated in the system's quantum jump trajectories, heralded by a reflected photon. Surprisingly, this maximally entangled state survives the strong coherent state input---a classical state that overwhelms the system. This simple method to generate maximally entangled states using classical coherent light and photon detection may, since our qubits are in a 1d continuum, find application as a building block of quantum networks. 
\end{abstract}
\maketitle

Quantum entanglement between two qubits is essential for quantum computing and indeed for quantum information processing more generally \cite{NielsenBook2010}. Bell states, which are maximally entangled two-qubit states, have perfect quantum correlations and are therefore especially important. The most common way to generate Bell states is to measure a joint property of two components and has been realized in several systems including, for example, 
trapped atoms, NV centers, quantum dots, and superconducting qubits 
(for reviews see \cite{Wendin_QinfProcess_review17,Brunner_Bell_RMP14,Aolita_OpenEntangle_RPP15}). 
Finding a variety of ways of making Bell states, particularly ones that use different resources, is important in advancing quantum information in new directions.
Since it is natural to suppose that classical resources decrease the coherence needed for entanglement, it is particularly interesting to produce Bell states using classical resources while reducing the quantum input to a minimum. 

A new platform named waveguide QED has recently been realized in which qubits strongly couple to photons confined in a one-dimensional (1d) waveguide \cite{LodahlRMP15, RoyRMP2016, NohRPP16, LiaoPhyScr16, GuPR17}. This platform has potential applications in integrating quantum components into complex systems, such as quantum networks \cite{CiracPRL97, KimbleNature2008}. 
In this work, we introduce a novel way of generating a Bell state of two qubits coupled to a 1d waveguide: \emph{classical} light plus photon detection leads to entanglement generation heralded by a reflected photon.
Previous results concerning entanglement in waveguide QED \cite{FicekPhyRep2002, GonzalezTudelaPRL2011, MartinCarnoPRB2011, CiccarelloPRA12, GonzalezBallesteroNJP2013, ZhengPRL2013, GonzalezBallesteroPRA2014, GangarajOE2015, Gonzalez-BallesteroPRB2015, LiaoPRA2015, PichlerPRA15, ChenPRA2016, FacchiPRA2016, HuOE2016, MirzaPRA2016, FacchiJPC2018}  have shown through analysis of the concurrence, entangled state population, or scattered wavefunction that a degree of entanglement between qubits can be generated using the effective interactions mediated by the waveguide. We show that under continuous monitoring, 
\emph{maximal entanglement} can be generated using the strong interference of photons in 1d and photon detection. This maximally entangled state is \emph{heralded} by detection of a reflected photon, which makes it attractive for potential applications. 

The driving in our system is a strong coherent state---a classical state that overwhelms the whole system. But surprisingly the Bell state survives this classical component.  What is more surprising and intriguing is that the steady state of the qubits is a trivial product state, which has no coherence or concurrence. The continuous monitoring unravels this trivial state such that its trajectories are non-trivial. This ``magical'' unravelling provides a particularly sharp illustration of the significance of the information gained about quantum systems by measurement, which has wide-reaching implications for advancing the understanding of quantum information and open quantum system.

\begin{figure}[tb]
\includegraphics[scale=0.45]{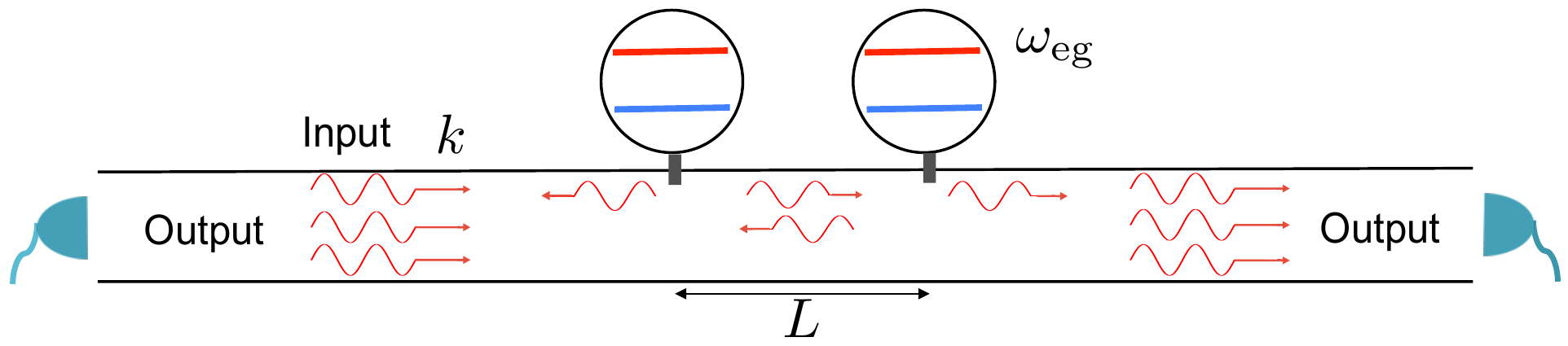}
\caption{Schematic of two qubits coupled to a 1d waveguide. We have a right-going coherent state as input from the left end. Transmitted and reflected photons are measured using photon counting detection at the right and left end respectively.}
\label{WQED}
\end{figure}

{\it Seemingly trivial steady state.--} The system we want to study, shown in Fig.\,\ref{WQED}, consists of two identical qubits coupled to a 1d waveguide under resonant driving by a coherent state. The input coherent state $\ket{\alpha}$ has frequency $k$, and the qubits with frequency $\omega_\textrm{eg}\!=\!k$ and raising (lowering) operators $\sigma_{i}^{\pm}$ ($i\!=\!1,2$) are separated by distance $L$. After tracing out the waveguide degrees of freedom making the Markov and rotating wave approximations, the two qubits can be described by a master equation of Lindblad form (see, e.g., \cite{FicekPhyRep2002, GonzalezTudelaPRL2011, ZhengPRL2013, LalumierePRA2013, ZhangPRA2018})
\begin{equation}\label{MEtrivial}
\frac{d}{dt} \rho= i \big[ \rho, H_\textrm{d}+H_\textrm{qq} \big] + \sum_{i,j=1,2} \Gamma_{ij} \big(\sigma^{-}_{i} \rho \sigma^{+}_{j} - \frac{1}{2} \{ \rho, \sigma^{+}_{i} \sigma^{-}_{j} \} \big).
\end{equation} 
The coherent evolution here has two parts, one describing the drive $H_\textrm{d} \!=\! g \alpha ( \sigma^{+}_{1} \!+\! \sigma_{2}^{+} e^{ikL}) \!+\! {\text{h.c.}}$ with coupling strength $g$, and the other $H_\textrm{qq} \!=\! \Omega ( \sigma^{+}_1 \sigma^{-}_2 \!+\! \sigma^{+}_2 \sigma^{-}_1 )$ describing a waveguide-mediated qubit-qubit interaction of strength $\Omega\!=\!2\pi g^2 \sin(k L)$.
In the incoherent Lindblad part, the individual decay rate of each qubit is $\Gamma_{11}\!=\!\Gamma_{22} \!\equiv\! \Gamma\!=\!4\pi g^2 $, and the cooperative decay $\Gamma_{12}\!=\!\Gamma_{21}\!=\!4\pi g^2 \cos{(kL)}$ is a waveguide-mediated incoherent coupling. 
The validity of the rotating wave and Markov approximations requires $\Gamma \!\ll\! \omega_\textrm{eg}$ and $\Gamma L \!\ll\! 1$; thus, $kL\sim1$ is clearly in the regime of validity.  



In the strong driving limit $\alpha \!\gg\! g$ 
(driving power $\gg \Gamma$), by letting $d \rho/dt \!=\! 0$ we obtain a trivial steady state in which the density matrix is an identity matrix. We consider $kL \!\neq\! n\pi$ where $n$ is an integer, in which case the steady state
$\rho_{\infty} =  ( \ket{ee}\!\bra{ee} + \ket{eg}\!\bra{eg} + \ket{ge}\!\bra{ge} + \ket{gg}\!\bra{gg} )/4$ is an identity matrix in the space spanned by 
$\{\ket{ee}, \ket{eg}, \ket{ge},  \ket{gg}\}$. 
[For $k L \!=\! n \pi$ where $n$ is an even (odd) integer, the steady state starting from the ground state is an identity matrix in the space spanned by $\{ \ket{ee}, \ket{gg}, \ket{\textrm{S}} (\ket{\textrm{A}}) \}$ where the symmetric and antisymmetric states are $\ket{\textrm{S}} (\ket{\textrm{A}}) \equiv (\ket{eg} \pm \ket{ge})/\sqrt{2}$.] 
This density matrix can be written simply as  
$\rho_{\infty} = ( \mathbbm{1}_{1} \otimes \mathbbm{1}_{2} )/4$ where $\mathbbm{1}_{i}$ is the identity matrix in the Hilbert space of $i$-th qubit. Therefore, the steady state has {\it no entanglement} (concurrence $\mathcal{C}=0$ \cite{WoottersPRL1998}) since it can be written as a product state and {\it no coherence} since there is no off-diagonal element. The qubit-qubit interaction mediated by the waveguide usually exploited to generate entanglement (see, e.g., \cite{GonzalezTudelaPRL2011}) is completely washed out by the classical driving and dissipation. However, the system's trajectories can be nontrivial, as we now show.


{\it Entanglement within trajectories.--} Our description in terms of a master equation is similar to that used for open quantum systems \cite{BreuerBookOQS2002}. In that context, the interaction between system and environment typically generates entanglement between them, and then a trace over the environmental degrees of freedom yields a mixed state for the system. During the partial trace, some information is lost as attested by the nonzero von Neumann entropy of the mixed state. However, under continuous monitoring, a mixed state can be unraveled as an ensemble of pure states (quantum trajectories) \cite{GardinerBook2004, CarmichaelBook2008, WisemanBook2014}. Unlike the mixed state, this ensemble gives a complete description of the open quantum system under continuous monitoring. 

Within the quantum trajectory description, mixed state entanglement can be defined without ambiguity as the average of pure state entanglement as follows \cite{NhaPRL2004}. Denote the ensemble of trajectories by $\{\sqrt{p_{i}} \ket{\psi_{i}}\}$, where $p_{i}$ is the probability of trajectory $\ket{\psi_{i}}$ being detected, and form $\rho=\sum_{i} p_{i} \ket{\psi_{i}} \bra{\psi_{i}}$. If we divide the open system into subsystems A and B, the entanglement between A and B within the $i$-th trajectory is defined through the usual von Neumann entropy as $S_{i} = - \Tr ( \rho_{i}^{A} \log_{2} \rho_{i}^{A} )$ with $\rho_{i}^{A} = \Tr_{B} ( \ket{\psi_{i}} \bra{\psi_{i}} )$. The entanglement in the ensemble is defined naturally as the average, $\bar{S} \equiv \sum_{i} p_{i} S_{i}$. 

It has been shown that measuring different quantities leads to different amounts of entanglement by unraveling with different ensembles of trajectories \cite{NhaPRL2004,ViviescasPRL2010, VogelsbergerPRA2010, MascarenhasPRA2011,ChantasriPRX2016}. 
For example, the trivial steady state above, $\rho_{\infty} \!=\! (\mathbbm{1}_{1} \!\otimes\! \mathbbm{1}_{2})/4$, can be unraveled nontrivially as either the ensemble $\{ \frac{1}{2} \ket{\Phi^{+}} , \frac{1}{2} \ket{\Phi^{-}} , \frac{1}{2} \ket{\Psi^{+}}, \frac{1}{2} \ket{\Psi^{-}}\}$ or $\{ \frac{1}{2} \ket{gg}, \frac{1}{2} \ket{ee}, \frac{1}{2} \ket{\Psi^{+}}, \frac{1}{2} \ket{\Psi^{-}}\}$, where $\ket{\Phi^{\pm}} \!=\! \big(\ket{gg} \!\pm\! \ket{ee} \big)/\sqrt{2}$ and $\ket{\Psi^{\pm}} \!=\! \big(\ket{ge} \!\pm\! \ket{eg} \big)/\sqrt{2}$ are the four conventional Bell bases. The former ensemble yields $\bar{S} \!=\! 1$ while the latter gives $\bar{S} \!=\! 1/2$ even though they both produce the seemingly trivial mixed state $\rho_{\infty}$. 

{\it Waveguide mediated collective jumps.--} Returning to our system, we suppose that photon counting measurements are performed at both ends of the waveguide, as shown in Fig.\,\ref{WQED}. As shown in our previous work \cite{ZhangPRA2018}, the photon detections at the left and right end can be described as discrete changes (quantum jumps) of quantum trajectories through the jump operators $J_{\text{L}}^{-}$ and $J_{\text{R}}^{-}$ defined as
\begin{equation}\label{JO}
\begin{split}
J_{\text{L}}^{-}&\equiv \sqrt{2\pi} g ( \sigma^{-}_{1} + \sigma_{2}^{-} e^{i k L } ), \\
J_{\text{R}}^{-}&\equiv \sqrt{2\pi} g ( \sigma_{1}^{-}  +  \sigma_{2}^{-} e^{-i k L} ) +  i \frac{\alpha}{\sqrt{2\pi}}  . \\
\end{split}
\end{equation}
Note that $J_\textrm{R}^-$ incorporates interference between the driving field $\alpha$ and the qubit emission. 
The master equation for the two qubits, Eq.\,(\ref{MEtrivial}), can be rewritten in an equivalent form as
\begin{equation}  \label{jumpME}
\frac{d}{dt} \rho  = i \big[ \rho, H_\textrm{h} \big] + \sum_{i = R, L} J^{-}_{i} \rho J^{+}_{i} - \frac{1}{2} \big\{ \rho , J^{+}_{i} J^{-}_{i} \big\},
\end{equation}
where $H_\textrm{h} \!=\! H_\textrm{qq} + \frac{1}{2} g \alpha ( \sigma^{+}_{1} + \sigma_{2}^{+} e^{i k L}) + {\text{h.c.}}$ \cite{ZhangPRA2018}. 
Based on the jump operator (\ref{JO}) that corresponds to photon detection, 
 quantum jump formalism \cite{WisemanBook2014} then yields quantum trajectories described by the stochastic Schr\"odinger equation (SSE)
\begin{equation} \label{SSE}
\begin{split}
d \ket{\psi (t)} & = \sum_{i=\text{L}, \text{R}}dN_{i}(t) \Big( \frac{J_{i}^{-} }{\sqrt{ \braket{ J_{i}^{+} J_{i}^{-} } } } -1 \Big) \ket{\psi(t)} \\
&  + \Big( \frac{ (1 - i \,dt\, H_{\text{eff}})  }{ | (1 - i \,dt\, H_{\text{eff}}) \ket{\psi(t)} | } - 1 \Big) \ket{\psi(t)},
\end{split}
\end{equation}
where $dN_{\text{i}}(t) \!=\! 0, 1$ describes the stochastic process of a photon being detected with probability $\braket{dN_{i} (t)} = dt \braket{\psi(t) | J_{\text{i}}^{+} J_{\text{i}}^{-} | \psi(t)}$, $dt$ is the time step, and $H_{\text{eff}} \equiv H_\textrm{h} - \frac{1}{2} \sum_{i = R, L} J^{+}_{i} J^{-}_{i}$ is the non-Hermitian effective Hamiltonian describing the segments of continuous evolution. 

It is intriguing that the left jump operator here, $J_{\text{L}}^{-} \sim (\sigma_{1}^{-} + e^{i k L} \sigma_{2}^{-}$), can produce a jump  $J_{\text{L}}^{-} \ket{ee} \rightarrow (\ket{ge} + e^{i k L} \ket{eg})$ that yields a maximally entangled state. This derives from the fact that detection of a reflected photon necessarily comes from a coherent superposition of the emission from both qubits, i.e.\ $\ket{ee} \rightarrow \ket{ge}$ and $\ket{ee} \rightarrow \ket{eg}$. This route to entanglement generation is in the same spirit as the scheme proposed in \cite{CabrilloPRA1999}. Note the following two requirements. (i)~To realize this jump process, the jump must start from $\ket{ee}$ or superpositions of $\ket{ee}$ and eigenstates of $J_{\text{L}}^{-}$ with vanishing eigenvalues. (ii) To make this maximally entangled state available for exploitation, it must not be destroyed for some time by the dynamics, such as the continuous evolution or subsequent jumps. We now show that when $kL \!=\! (n \!+\! 1/2) \pi$ and the driving $\ket{\alpha}$ is strong, these two requirements can be met.


{\it Hybridizing jumps and state diffusion.--}  In the strong driving limit $\alpha /g \!\rightarrow\! \infty$, each right jump leads to an infinitesimal change of the wavefunction, since the right jump operator $J^{-}_{\text{R}}$ is dominated by the constant term. However, within a time step $dt$ there will be infinitely many right jumps due to the large photon flux given by the strong coherent state. Therefore, the quantum trajectory will be continuous, as in classic homodyne detection \cite{WisemanBook2014} when left jumps are absent and the photon current is measured. Then, the number of right jumps detected in a time step, denoted $dN_{\text{R}} (t)$, can be written as
\begin{equation}\label{dNR}
dN_{\text{R}} (t)  = \braket{ dN_{\text{R}} (t) } + \frac{|\alpha|}{\sqrt{2\pi}} d\xi(t),
\end{equation}
where $d\xi(t)$ is stochastic noise. Since the coherent state dominates the signal detected, Gaussian noise
with $\braket{d\xi(t)}=0$ and $\braket{d\xi(t)^2} = dt$ is a good approximation. 
By expanding in $1/|\alpha|$, the SSE Eq.\,(\ref{SSE}) is simplified to (for details see \cite{SupMat})
\begin{widetext}
\begin{equation}\label{QSD}
\begin{split}
d\ket{\tilde{\psi}(t)} = \,\,& dt \Big( -i ( g \alpha c^{+} + g \alpha^{*} c^{-} + H_{qq}) -i e^{-i\theta} \pi g^2 \braket{( i e^{i \theta} c^{+} - i e^{-i\theta} c^{-} ) } c^{-} - \pi g^2 c^{+} c^{-} - \frac{1}{2} J_{\text{L}}^{+} J_{\text{L}}^{-}  \Big) \ket{\tilde{\psi}(t)} \\
                                &+ d\xi(t) \Big(-i e^{-i \theta} \sqrt{2\pi}g c^{-}  \Big) \ket{\tilde{\psi}(t)} + dN_{\text{L}}(t) \Big( \frac{J_{\text{L}}^{-} }{\sqrt{ \braket{ J_{\text{L}}^{+} J_{\text{L}}^{-} } } } -1 \Big) \ket{\tilde{\psi}(t)},
\end{split}                      
\end{equation}
\end{widetext}
where $\ket{\tilde{\psi}}$ is an unnormalized wavefunction, 
$\alpha \!=\! | \alpha | e^{i \theta}$, 
$\braket{\cdot} \!=\! \braket{\psi | \cdot | \psi}$, and 
$c^{\pm} \!\equiv\! ( \sigma_{1}^{\pm} + e^{\pm i k L} \sigma_{2}^{\pm} )$ is the operator part of $J_{\text{R}}^{-}$ such that 
$J_{\text{R}}^{-}\!=\! \sqrt{2\pi} g  c^{-}  \!+  i \alpha  / \sqrt{2\pi}$. 
If the left jumps are dropped, note that this SSE becomes a quantum state diffusion equation with fluctuations given by a Weiner process $d \xi(t)$. 

{\it Heralded Bell state.--} We wish to focus on the case $kL \!=\! (n\!+\!1/2) \pi$, where $n$ is an even (odd) integer, and define two maximally entangled states $\ket{\pm i} \!\equiv\! ( \ket{ge} \!\pm\! i \ket{eg} )/\sqrt{2}$ (Bell states). Then, the operator $c^{-}$ ($J_{\text{L}}^{-}$) is a lowering operator in the space spanned by $\{ \ket{ee}, \ket{-i}, \ket{gg}\}$ while $J_{\text{L}}^{-}$ ($c^{-}$) is a lowering operator in the space spanned by $\{ \ket{ee}, \ket{+i}, \ket{gg}\}$. In the following, we let $kL \!=\! \pi/2$, i.e.\,the qubit separation is a quarter wavelength. For other even $n$, the conclusions are the same; for odd $n$, they hold upon switching the roles of $\ket{\pm i}$.

\begin{figure}[b]
\includegraphics[scale=0.4]{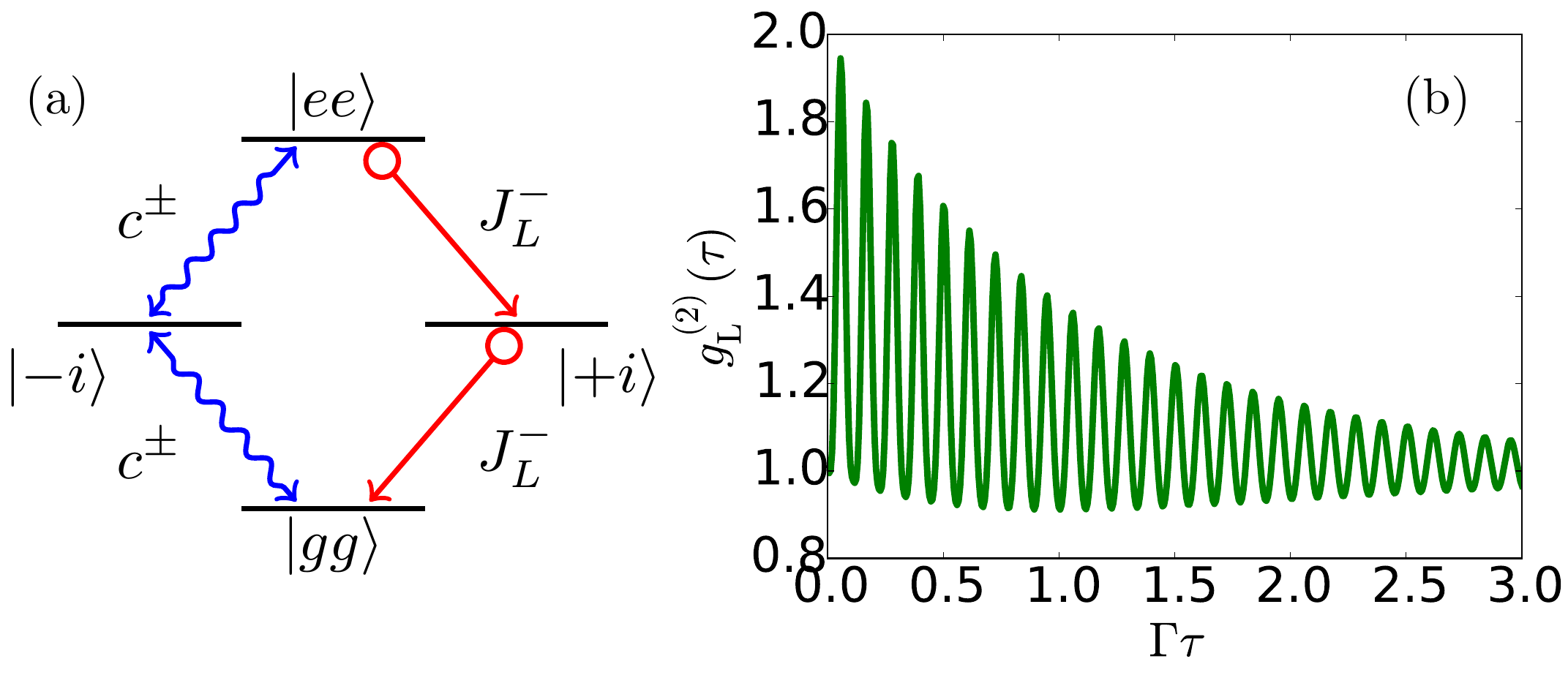} 
\caption{(a) Energy level diagram for $k L \!=\! \pi/2$. Red and blue arrows represent left jumps and driving, respectively, $\ket{\pm i} \!\equiv\! ( \ket{ge} \pm i \ket{eg} )/\sqrt{2}$, $J^{-}_{\text{L}}$ is the left jump operator, and $c^{\pm}$ comes from the right jump operator $J^{-}_{\text{R}}$. The effective qubit-qubit interaction $H_\textrm{qq}$ is suppressed by the strong driving. (b) Second order correlation function for the reflected photons calculated from input-output theory. (Parameters: $k L = \pi/ 2$, $\alpha=100$.)}\label{level}
\end{figure}

The energy level diagram for $kL \!=\! \pi/2$ is shown in Fig.\,\ref{level}(a). The quantum diffusion process given by the operator $c^{\pm}$  causes 
$\ket{gg} \!\leftrightarrow\! \ket{-i} \!\leftrightarrow\! \ket{ee}$, 
and the left jump process causes $\ket{ee} \!\rightarrow\! \ket{+i} \rightarrow\! \ket{gg}$. Thus, the two maximally entangled states $ \ket{\pm i}$ are dynamically separated. The ground state of the qubits $\ket{gg}$ will be driven to the excited state $\ket{ee}$, from which there is a finite probability for a left jump. In that case, the two qubits jump to the maximally entangled state $\ket{+i}$, while at the same time a left-going (reflected) photon is detected. The qubits will stay in $\ket{+i}$ until a second left jump occurs, taking the qubits back to $\ket{gg}$. The whole process then repeats. Thus, there are repeated windows of maximally entangled state $\ket{+i}$, whose lifetime is $1/\braket{+i | J_{\text{L}}^{+} J_{\text{L}}^{-} |+i} = 1/\Gamma$, each heralded by a reflected photon. 

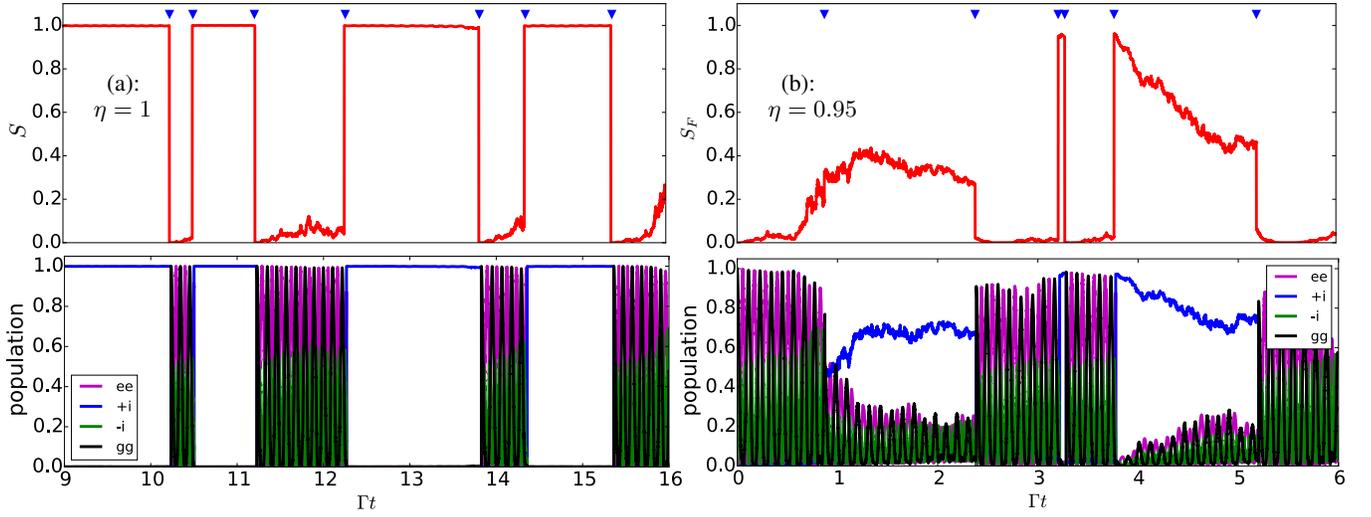
\begin{figure*}
\includegraphics[width=\textwidth]{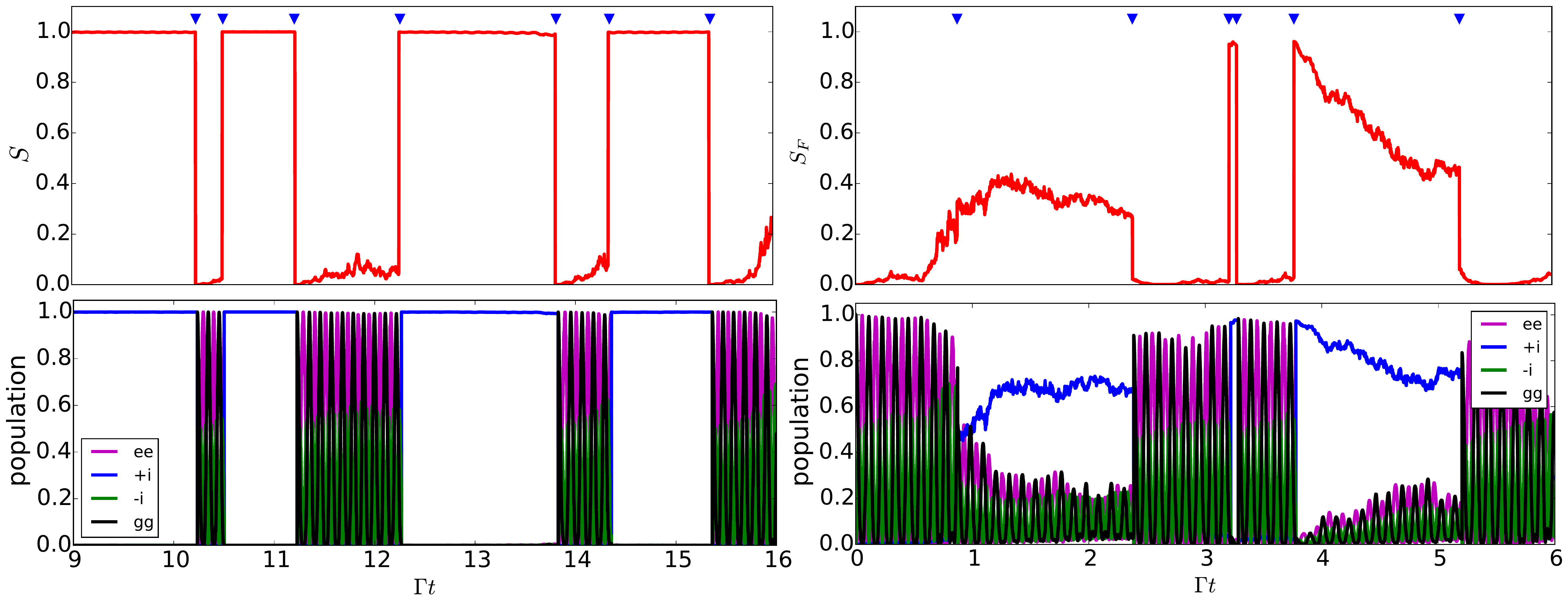} 
\put(-470,160){(a):}
\put(-475,150){$\eta =1$}
\put(-215,160){(b):}
\put(-220,150){$\eta =0.95$}
\caption{Example trajectories of entanglement (first row) and populations (second row) for (a) perfect photon detection and (b) lossy photon detection with efficiency $\eta_{\text{i=\text{L,R}}}\!=\!\eta\!=\!0.95$. The entanglement for pure states in (a) and mixed states in (b) is quantified using the von Neumann entropy $S$ and the entanglement of formation $S_{F}$, respectively. The times at which quantum jumps occur are marked with blue triangles. Longer trajectories, from $\Gamma \tau=0$ to $20$, are shown in \cite{SupMat}. (Parameters: $k L = \pi/ 2$, $\alpha=100$, qubits initially in the ground state $\ket{gg}$.) } \label{trajectory}
\end{figure*}

An example trajectory is shown in Fig.\,\ref{trajectory}(a) for $\alpha\!=\!100$.
There are clearly time windows of maximal entanglement, whose birth and death are heralded by the detection of reflected photons. The populations of the energy levels show that the qubits are in the $\ket{+i}$ state in the maximal entanglement windows and are dynamically decoupled from the other three levels in these windows. The small deviations from maximal entanglement that can be seen 
are due to the effective qubit-qubit interaction term $H_\textrm{qq}$ that exchanges excitations between the two qubits and so leads to the process $\ket{+i} \!\leftrightarrow\! \ket{-i}$. This term ($\sim\! g^2$) is suppressed by the strong driving term ($\sim\! g |\alpha|$) as shown in \cite{SupMat}, which is the reason why strong driving is needed. Outside the windows of maximal entanglement, the dynamics is dominated by Rabi oscillations in a three-level system with fluctuations coming from the Weiner process. 

This special dynamics is encoded in the behavior of the second-order correlation function $g^{(2)}_{\text{L}}(\tau)$ of the reflected light, shown in Fig.\,\ref{level}(b). $g^{(2)}_{\text{L}}$ starts at $1$ and then oscillates at the Rabi frequency with an envelope that decays in a time of order $\Gamma^{-1}$. It is bounded by $2$ and reaches maximal points when $\ket{gg}$ is driven to $\ket{ee}$ (see \cite{SupMat} for details).

When parameters are detuned from their ideal values (either $k$ or $L$), the dynamics becomes more complicated than shown in Fig.\,\ref{level}(a), with for instance a (weak) direct connection between the left and right sides. For small detuning, the dynamics will be qualitatively similar; we leave a quantitative study of these features to future work.

{\it Imperfect photon detection.--} To understand the role and importance of the information gained by observing a quantum system, we introduce information loss through imperfect photon detection. The effect of such loss is modeled using the jump operators $\sqrt{\eta_i} J^{-}_{i} $, where $i \!=\! \text{R}, \text{L}$ and $\eta_i < 1$ is the efficiency of photon detection \cite{WisemanBook2014}. 
Then the SSE \eqref{QSD} becomes a stochastic master equation (SME) (for details see \cite{SupMat}),
\begin{widetext}
\begin{equation}\label{SME}
\begin{split}
d \tilde{\rho_{s}}(t) = \,\,& dt \Big( i [ \tilde{\rho_{s}}, H_{qq} + g \alpha^{*} c^{-} + {\text{h.c.}} ] + (1 - \eta_{\text{L}} ) J_{\text{L}}^{-} \tilde{\rho_{s}} J_{\text{L}}^{+} + 2 \pi g^2 c^{-} \tilde{\rho_{s}} c^{+}   - \frac{1}{2} \{\tilde{\rho_{s}} , J_{\text{L}}^{+} J_{\text{L}}^{-} + 2 \pi g^2 c^{+} c^{-} \} \Big) \\
                                &+ d\xi(t) \sqrt{\eta_{\text{R}}} \Big(-i e^{-i \theta} \sqrt{2\pi}g c^{-}  \tilde{\rho_{s}}  + {\text{h.c.}}  \Big)  + dN_{\text{L}}(t) \Big( \frac{ J_{\text{L}}^{-}  \tilde{\rho_{s}} J_{\text{L}}^{+} }{ \Tr[ J_{\text{L}}^{-}  \tilde{\rho_{s}} J_{\text{L}}^{+} ] } -\tilde{\rho_{s}} \Big),
\end{split}                      
\end{equation}
\end{widetext}
for  trajectories of mixed states $\tilde{\rho_{s}}$ 
\footnote{Note that for perfect photon detection ($\eta =1$), when unnormalized, our notation $\tilde{\rho_{s}}  =  c(t) \ket{\tilde{\psi}} \bra{\tilde{\psi}}$, where $c(t) \neq 1$ due to their different normalization factors. After normalization, $\rho_{s} = \ket{\psi} \bra{\psi}$.}
due to loss of information about the system. The probability of photon detection now becomes $\braket{dN_{\text{L}}} \!=\! \eta_{\text{L}} dt  \Tr[ \rho_{s} J_{\text{L}}^{+} J_{\text{L}}^{-}] $ in terms of the normalized density matrix $\rho_{s} \!=\! \tilde{\rho_{s}}/\Tr [\tilde{\rho_{s}}]$. Other information loss mechanisms, such as the coupling of the qubits to channels other than the waveguide, can be taken into account by simply adding additional Lindbladian dissipators to Eq.\,\eqref{SME}; however, this will produce no  qualitative change in our results and so is left to the interested reader.

We quantify the entanglement for each mixed trajectory using the entanglement of formation $S_{\text{F}}$ \cite{WoottersPRL1998}. 
To define $S_{\text{F}}$, consider a ``purification'' of a mixed state, by which is meant a pure state of the system plus environment that yields the known mixed state through partial trace. 
The entanglement entropy of a purification is simply that of the two qubits, $\bar{S}$, conditioned on measurement of the environment (photon detection here). 
The entanglement of formation $S_{\text{F}}$ is the minimum entanglement entropy for all possible purifications of a mixed state, and so  gives a lower bound on the entanglement contained in a mixed trajectory. 
A subtle point should be emphasized here: information gained about a quantum system constrains possible purifications and therefore gives a different lower bound. 
For our system (assume $\eta_{i}\!=\!\eta$ for now), for example, if $\eta=0$, i.e.\ no photons are measured so no information is gained, Eq.\,\eqref{SME} becomes Eq.\,\eqref{jumpME} whose steady state is $(\mathbbm{1}_{1} \otimes \mathbbm{1}_{2})/4$ and $S_{F}=0$. As $\eta$ increases, more information is gained and the number of possible purifications decreases. When $\eta\!=\!1$, Eq.\,\eqref{SME} becomes Eq.\,\eqref{QSD}, which becomes the only way to purify given the physical setup. 

An example trajectory for $\eta\!=\!0.95$ is shown in Fig.\,\ref{trajectory}(b). As can be seen, the information loss leads to very different behavior. In the first window, the entanglement $S_{F}$ and the $\ket{+i}$ population do not jump up to $1$ as for perfect photon detection. This is because there is a possibility that photons have been emitted without being detected, as shown by the term $(1 - \eta_{\text{L}} ) J_{\text{L}}^{-} \tilde{\rho_{s}} J_{\text{L}}^{+} $ in Eq.\,\eqref{SME}, which makes the trajectory be in the space spanned by all four energy levels. When a photon is detected, the trajectory is projected to a space spanned by $\{ \ket{+i}, \ket{gg} \}$ through processes $\ket{ee} \rightarrow \ket{+i}$ and $\ket{+i} \rightarrow \ket{gg}$. In the third window, although the qubits jump to $\ket{+i}$, its population keeps decreasing with time. This is because of the undetected decaying process $\ket{+i} \rightarrow \ket{gg}$. 

{\it Only one detector needed.--} Even though the scheme proposed here is not robust against photon detection loss at the left end, it works independently of the photon detection efficiency at the right end. It can be seen in Eq.\,\eqref{SME} that, as long as $\eta_{\text{L}}\!=\!1$, the continuous part describes time evolution of a mixed state in the space spanned by $\{ \ket{ee}, \ket{-i}, \ket{gg}\}$ and the jump part still describes detection of reflected photons, which project the $\ket{ee}\bra{ee}$ component onto a pure state $\ket{+i}$ as shown in Fig.\,\ref{level}(a) \cite{SupMat}. That is, the scheme still works even without photon detection at the right end ($\eta_{\text{R}}=0$). 


{\it Conclusion and outlook.--} In summary, we have shown that for two qubits coupled to a waveguide separated by $(n/2+1/4)$ wavelengths, a heralded Bell state can be generated using classical driving and photon counting detection. Although the steady state is a trivial product state, the continuous monitoring unravels the master equation such that a Bell state is dynamically decoupled from the other three levels during the continuous part of the evolution. Discrete jumps, heralded by detections of reflected photons, project the wavefunction onto the Bell state. This physical example that non-entangled mixed states can have entangled trajectories calls for careful usage of commonly used entanglement measures, such as concurrence, especially when measurement is present. Since the qubits are already in the continuum and coupled to itinerant photons, the method presented here will have particular application in integrating quantum components into complex systems  \cite{CiracPRL97, KimbleNature2008}.

The importance of the information gained by observing a quantum system is shown by introducing information loss caused by imperfect photon detections. A small information loss causes the quantum entanglement to behave very differently. This implies that methods to stabilize the Bell state, such as bath engineering \cite{KapitQST2017}, are needed in applications. 

In this work, the Markov approximation has been applied, which is valid when the qubit separation is not too large. It will be interesting to explore in the future the effects caused by time delayed feedback in the non-Markovian regime \cite{GonzalezBallesteroNJP2013, ZhengPRL2013, LaaksoPRL14, GrimsmoPRL2015, FangPRA15, *FangPRA15err, PichlerPRL2016, GuimondQST17, FangNJP18, CalajoPRL2019}, which is important for the generation of remote entanglement between qubits. 

\begin{acknowledgements}
We thank T.\ Barthel and I.\ Marvian for helpful conversations. 
This work was supported in part by U.S.\,DOE, Office of Science, Division of Materials Sciences and Engineering, under Grant No.\,{DE-SC0005237}.
\end{acknowledgements}

\bibliography{Entangle,WQED_2018-07,BibFootnotes}


\newpage

\widetext

\clearpage

\setcounter{equation}{0}
\setcounter{figure}{0}
\setcounter{table}{0}
\setcounter{page}{1}
\makeatletter
\renewcommand{\theequation}{S\arabic{equation}}
\renewcommand{\thefigure}{S\arabic{figure}}

\begin{center}

{\large\bf Supplemental Material for ``Heralded Bell State of Dissipative Qubits Using Classical Light in a Waveguide''}

\vspace{0.5cm}
 
Xin H. H. Zhang and Harold U. Baranger

{\it Department of Physics, Duke University, P.\,O.\,Box 90305, Durham, NC 27708-0305, USA}
\end{center}

In this Supplemental Material, we present  
(i)  derivation of the stochastic Schr\"odinger equation (SSE) \eqref{QSD},
(ii) analysis of the qubit-qubit interaction
(iii) analysis of the second-order correlation function shown in Fig.\ref{level}(b),
(iv) derivation of the stochastic master equation (SME) \eqref{SME},
(v) analysis of the single detector case
and (vi) some trajectories to complement those shown in Fig.\,\ref{trajectory}.

\section{(i) Derivation of Eq.\,\eqref{QSD}, the stochastic Schr{\"o}dinger equation (SSE)}

In the strong driving limit $\alpha \rightarrow \infty$, 
\begin{equation}
J_{\text{R}}^{-}=\sqrt{2\pi} g c^{-}  +  i \alpha  / \sqrt{2\pi} \;\propto\; 1 + \frac{2 \pi g}{i \alpha} c^{-},
\end{equation} 
where $c^{-} \!\equiv\! ( \sigma_{1}^{-} + e^{- i k L} \sigma_{2}^{-} )$.
This leads to an infinitesimal change of the wavefunction for each right jump. On the other hand, there are $O(dt |\alpha|^2)$ jumps, which is large for a finite time bin $dt$. The situation is then similar to quantum state diffusion and homodyne detection. The number of right jumps $d N_{\text{R}}(t)$ in \eqref{SSE}, which is dominated by the strong coherent state, can be represented as a Gaussian noise as shown in \eqref{dNR}. Expanding the RHS of \eqref{SSE} over $1/|\alpha|$ gives the quantum state diffusion equation \eqref{QSD} for unnormalized wavefunction $\ket{\tilde{\psi}}$, where terms like $(\text{complex number}) \times \ket{\psi}$ have been omitted since they can be retrieved during normalization.

\section{(ii) Analysis of the Qubit-Qubit Interaction}

In this section, we want to show that the qubit-qubit interaction $H_{qq}$ can be ignored when the driving is strong. When there are no jumps ($dN_{\text{L}}=0$), the dynamics in Eq.\,\eqref{QSD} can be captured by considering a simplified Hamiltonian
\begin{equation}
H^{\prime} = ( H_{D} + H_{qq}) -i \pi g^2 c^{+} c^{-} -i \frac{1}{2} J_{\text{L}}^{+} J_{\text{L}}^{-},
\end{equation}
where $H_{D} = g \alpha (c^{+} + c^{-}) = g \alpha( \sigma^{x}_{1} - \sigma^{y}_{2})$, $H_\textrm{qq} \!=\! 2 \pi g^2 ( \sigma^{+}_1 \sigma^{-}_2 \!+\! \sigma^{+}_2 \sigma^{-}_1 ) = \pi g^2 (\sigma_{1}^{x}\sigma_{2}^{x} + \sigma_{1}^{y}\sigma_{2}^{y})$ and $\alpha$ has been assumed to be real without loss of generality. Then in an interaction picture with respect to the driving term $H_{D}$, 
\begin{equation}
H_{qq} (t) = e^{iH_{D}t} H_{qq} e^{-iH_{D}t} = \pi g^2 \Big\{\sigma_{1}^{x} \big( \sigma_{2}^{x} \cos(2g\alpha t) + \sin(2g\alpha t) \sigma_{2}^{z} \big)+ \big(\sigma_{1}^{y} \cos(2g\alpha t) +\sigma_{1}^{z} \sin(2g\alpha t) \big) \sigma_{2}^{y}\Big\},
\end{equation}
which is clearly a fast rotating term when $\alpha \gg g$ and can therefore be ignored. This can be understood intuitively by noticing that $ e^{-i H_{qq} \Delta t}\ket{+i} = \cos(2\pi g^2 \Delta t) \ket{+i} + \sin(2\pi g^2 \Delta t) \ket{-i}$, which means that the population of $\ket{+i}$ decreases slowly when $\Delta t$ is small. The strong driving term $H_{D}$ then quickly drives the $\ket{-i}$ population away and starts a fast Rabi oscillation in $\{ \ket{gg}, \ket{-i}, \ket{ee} \}$. This means that in the next step the $\ket{-i}$ level is effectively empty again, which leads to a slow decrease of the $\ket{+i}$ population.

\section{(iii) Analysis of the Second-Order Correlation Function shown in Fig.\ref{level}(b)}

\begin{figure}[h!]
\includegraphics[scale=1.4]{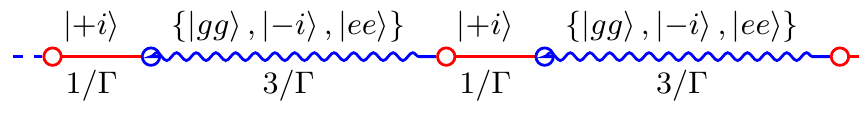} 
\caption{Schematic illustration of photon detection events for the reflected light when $k L = \pi/ 2$ and $\alpha \rightarrow \infty$. Circles denote the clicks of photon detection. Red straight lines denote the periods when the qubits are in the $\ket{+i}$ state and blue curly lines denote the periods when the qubits are undergoing Rabi oscillation between the three levels $\{ \ket{gg}, \ket{-i}, \ket{ee} \}$ [the left half of Fig.\,\ref{level}(a)]. $1/\Gamma$ and $3/\Gamma$ are their average lifetime respectively.}\label{Flux}
\end{figure}

The second order correlation function can be calculated using input-output theory (see, e.g.\,\cite{ZhangPRA2018} and references therein). Here we present an intuitive explanation for better understanding of the system's dynamics. The second order correlation function can be understood in terms of a conditional probability using the relation $g^{(2)}(\tau) = P( t+\tau | t) / P(t+\tau)$, where $P( t+\tau | t)$ is the probability density of detecting a photon at time $t +\tau$ conditioned on a photon detection at $t$ while $P( t+\tau)$ is the probability density of detecting a photon at time $t +\tau$ without any previous knowledge \cite{ZhangPRA2018}. In the steady state, which is the situation considered here, there is no $t$ dependence, and $P( t+\tau)$ is simply the photon flux. 

A detection of a reflected photon at $t$ can result from two equally probable processes: (i) $\ket{ee} \rightarrow \ket{+i}$ or (ii) $\ket{+i} \rightarrow \ket{gg}$. The former process will lead to a subsequent decay of $\ket{+i}$, which gives a probability density $\Gamma$ at $t^{+}$ for detection of a second photon. In the latter process, the qubits are projected onto $\ket{gg}$ at $t^{+}$, which cannot decay anymore. 

The system undergoes Rabi oscillations between the three levels $\{ \ket{gg}, \ket{-i}, \ket{ee} \}$ as shown in the left half of Fig.\,\ref{level}(a). Since the system is oscillating between three levels and only $\ket{ee}$ can decay, the lifetime of this oscillating state is $3 \times 1/\Gamma$. A schematic illustration of photon detection events for the reflected light is shown in Fig.\,\ref{Flux}. The average photon flux is $2$ photons in time $4/\Gamma$; thus, the probability density $P(t+\tau)$ is simply $\Gamma/2$. 

Because at $t+0^{+}$ process (i) gives a decay rate $\Gamma$ and process (ii) gives decay rate $0$, we have $P(t+0^{+} | t) = 1/2 \times \Gamma + 1/2 \times 0 = \Gamma/2$. Therefore, $g^{(2)}_{\text{L}} (0) = 1$. This can also be seen mathematically by putting $\rho_{\infty} = 1/4$ into the definition of $g^{(2)}(0) = \braket{J_{\text{L}}^{+} J_{\text{L}}^{+} J_{\text{L}}^{-} J_{\text{L}}^{-}} / \braket{J_{\text{L}}^{+} J_{\text{L}}^{-}}^2$. To find  $g^{(2)}_{\text{L}} (\tau)$, note that the population of $\ket{ee}$ starts oscillating after each type (ii) process and yields a maximal decay rate $\Gamma$ when $\ket{ee}$ is fully occupied. Therefore $P(t+\tau | t)$ can reach a value no larger than $1/2 \times \Gamma + 1/2 \times \Gamma = \Gamma$. In fact, $P(t+\tau | t)$ will be smaller as $\tau$ increases due to the decay of the $\ket{+i}$ population after process (i). These results for $P(t+\tau | t)$ imply, therefore, that $g^{(2)}_{\text{L}} (\tau)$ is bounded by $2$ and oscillates at the Rabi frequency with an envelope that decays in a time of order $\Gamma^{-1}$.

\section{(iv) Derivation of Eq.\,\eqref{SME}, the stochastic master equation (SME)}

For a mixed state $\rho$, after a photon detection given by $J^{-}_{i}$, 
\begin{equation}
\rho \rightarrow \frac{J^{-}_{i} \rho J^{+}_{i}  }{ \Tr[J^{-}_{i} \rho J^{+}_{i}]}.
\end{equation}
For imperfect photon detection, using the jump operator $\sqrt{\eta_i} J^{-}_{i}$, the master equation \eqref{jumpME} can now be written as 
\begin{equation} \label{SMEjump}
\begin{split}
\frac{d}{dt} \rho  &= i \big[ \rho, H_{h} \big] + \sum_{i = R, L} \Big( (1 - \eta_i ) J^{-}_{i} \rho J^{+}_{i} - \frac{1}{2} \big\{ \rho , J^{+}_{i} J^{-}_{i} \big\}  \Big) + \sum_{i = R, L} \eta_i  J^{-}_{i} \rho J^{+}_{i} \\
      			       &\equiv \mathcal{D}_{\text{eff}}[\rho] + \sum_{i = R, L} \eta_i  J^{-}_{i} \rho J^{+}_{i},
\end{split}
\end{equation}
which can be unravelled as a stochastic master equation:
\begin{equation} \label{SMEDiffusive}
d \rho_{s} = \Big( \frac{ \rho_{s} + dt \mathcal{D}_{\text{eff}}[\rho_{s}] } {\Tr\big[ \rho_{s} + dt \mathcal{D}_{\text{eff}}[\rho_{s}]  \big] } - \rho_{s} \Big) + \sum_{i = R, L} dN_{i}  \Big(  \frac{J^{-}_{i} \rho_{s} J^{+}_{i}  }{ \Tr[J^{-}_{i} \rho_{s} J^{+}_{i}]} - \rho_{s} \Big),
\end{equation}
where $dN_{\text{i}} =0, 1$ and $\braket{dN_{i}} = \eta_i dt  \Tr[ \rho_{s} J_{\text{i}}^{+} J_{\text{i}}^{-}] $.

In the strong driving limit, $d N_{\text{R}}$ can be treated as a Gaussian noise source,
\begin{equation}
dN_{\text{R}}  = \braket{ dN_{\text{R}} } + \sqrt{\eta_{\text{R}}} \frac{|\alpha|}{\sqrt{2\pi}} d\xi(t).
\end{equation}
Expanding \eqref{SMEDiffusive} over $1/|\alpha|$ gives the unnormalized stochastic master equation shown in \eqref{SME}. For $\eta = 1$, after normalization, agreement between \eqref{QSD} and \eqref{SME} can be found by noticing that $d\xi ^2 = dt$.

\newpage
\section{(v) Analysis of the single detector case}

\begin{figure}[h]
	\includegraphics[width=\textwidth]{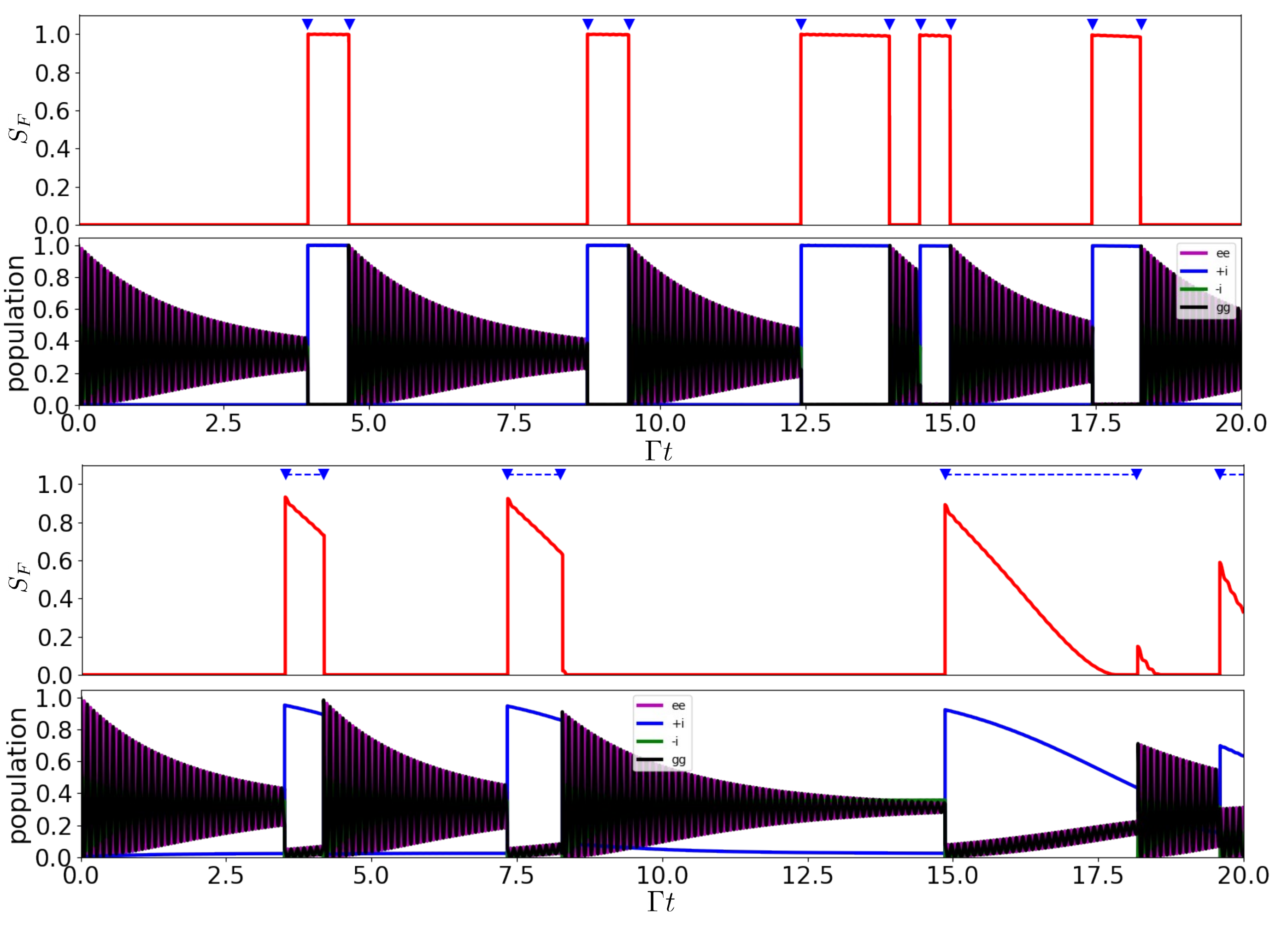} 
	\put(-470,360){(a):}
	\put(-470,350){$\eta_{\text{L}}=1$}
	\put(-470,340){$\eta_{\text{R}}=0$}
	\put(-470,160){(b):}
	\put(-470,150){$\eta_{\text{L}}=0.95$}
	\put(-470,140){$\eta_{\text{R}}=0$}
	\caption{Example trajectories of entanglement (first and third row) and populations (second and fourth rows) for (a) perfect photon detection ($\eta_{\text{L}}\!=\!1$) and (b) lossy photon detection ($\eta_{\text{L}}\!=\!0.95$) when only the reflected photons are detected i.e.\,$\eta_{\text{R}}=0$. The entanglement is quantified with the entanglement of formation $S_{F}$. The times at which quantum jumps occur are marked with blue triangles. (Parameters: $k L = \pi/ 2$, $\alpha=100$, qubits initially in the ground state $\ket{gg}$.) }
	\label{SingleDetector}
\end{figure}

To show that our scheme works independently of the detection efficiency of transmitted photons, we present here the extreme case where there is only one detector for the reflected photons, which can be described by setting $\eta_{\text{R}} = 0$ in Eq.\,\eqref{SME}. Example trajectories for $\eta_{\text{L}}=1$ and $\eta_{\text{L}}=0.95$ are shown in Fig.\,\ref{SingleDetector}. It can be seen that the maximally entangled state can be generated by detecting the reflected photons only. When there is photon loss ($\eta_{\text{L}}<1$), the loss of information leads to non-maximal entanglement like the two detector case discussed in the main text. Note that it is less noisy here because there are no fluctuations coming from the detection of transmitted photons.

\newpage
\section{(vi) Example Trajectories from $\Gamma t=0$ to $\Gamma t =20$ }

\begin{figure}[h]
\includegraphics[width=\textwidth]{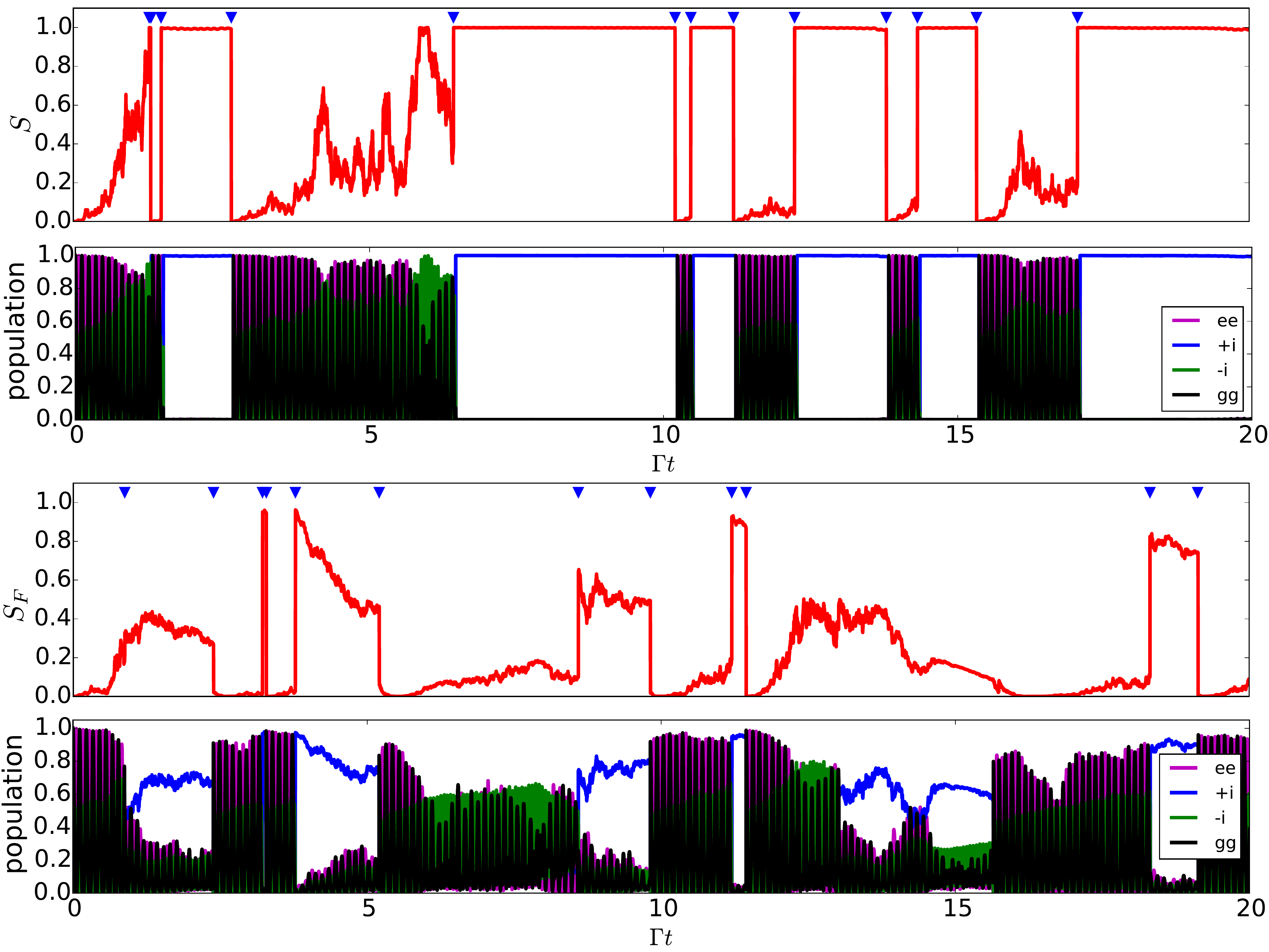} 
\put(-470,360){(a):}
\put(-475,350){$\eta=1$}
\put(-470,160){(b):}
\put(-475,150){$\eta=0.95$}
\caption{Example trajectories of entanglement (first and third row) and populations (second and fourth rows) for (a) perfect photon detection and (b) lossy photon detection with efficiency $\eta_{i=\text{L,R}}\!=\!0.95$. The entanglement for pure states in (a) and mixed states in (b) is quantified using the von Neumann entropy $S$ and the entanglement of formation $S_{F}$, respectively. The times at which quantum jumps occur are marked with blue triangles. (Parameters: $k L = \pi/ 2$, $\alpha=100$, qubits initially in the ground state $\ket{gg}$.) }
\label{LongTraj}
\end{figure}

\end{document}